\documentclass{article}%
\usepackage{aip}
\usepackage{amsmath}
\usepackage{graphicx}
\usepackage{amsfonts}
\usepackage{amssymb}%
\providecommand{\U}[1]{\protect\rule{.1in}{.1in}}
\providecommand{\U}[1]{\protect\rule{.1in}{.1in}}
\providecommand{\U}[1]{\protect\rule{.1in}{.1in}}
\providecommand{\U}[1]{\protect\rule{.1in}{.1in}}
\providecommand{\U}[1]{\protect\rule{.1in}{.1in}}
\providecommand{\U}[1]{\protect\rule{.1in}{.1in}}

\begin{document}

\title{Consistency Problem of the Solutions of the Space Fractional Schr\"{o}dinger Equation}
\author{Sel\c{c}uk \c{S}. Bayin\\Middle East Technical University\\Institute of Applied Mathematics\\Ankara TURKEY 06800}
\date{\today}
\maketitle

\begin{abstract}
Recently, consistency of the infinite square well solution of the space
fractional Schr\"{o}dinger equation has been the subject of some controversy.
In [J. Math. Phys. 54, 014101 (2013)], Hawkins and Schwarz objected to the way
certain integrals are evaluated to show the consistency of the infinite square
well solutions of the space fractional Schr\"{o}dinger equation [J. Math.
Phys. 53, 042105 (2012); J. Math. Phys. 53, 084101 (2012)]. Here, we show for
general $n$ that as far as the integral representation of the solution in the
momentum space is concerned, there is no inconsistency. To pinpoint the source
of a possible inconsistency, we also scrutinize the different representations
of the Riesz derivative that plays a central role in this controversy and show
that they all have the same Fourier transform, when evaluated with consistent assumptions.


\end{abstract}

\section{Introduction}

Fractional calculus is an effective tool in the study of non local and memory
effects in physics. Its successful application to anomalous diffusion was
immediately followed by other examples in classical physics [1-4]. The first
application of fractional calculus to quantum mechanics was given by Laskin in
terms of the fractional Riesz derivative as the space fractional
Schr\"{o}dinger equation [5]. Laskin's space fractional quantum mechanics is
intriguing since it follows from the Feynman's path integral formulation of
quantum mechanics over L\'{e}vy paths. One of the first solutions of this
theory was given by Laskin for the infinite well problem [5-8]. Despite its
simplicity, the infinite well problem is very important since it is the
prototype of a quantum detector with internal degrees of freedom. In 2010,
Jeng et. al. [9] argued that the solutions obtained for the space fractional
Schr\"{o}dinger equation in a piecewise fashion are not valid. Their argument
was based on a contradiction they think exists in the ground state wave
function of the infinite square well problem. In [10, 11] we have shown that
an exact treatment of the integral that lead them to inconsistency proves
otherwise. However, in a recent comment, Hawkins and Schwarz point to a
possible problem in the proof regarding the analyticity of the relevant
integrals [12].

In Sections II and III we present details of the treatment of the relevant
integrals and show for general $n$ that there is no inconsistency. Recently,
Dong [13] obtained the wave function for the infinite square well problem by
using path integrals over L\'{e}vy paths and confirmed the solution given by
Laskin [5-8].

However, Luchko analyzed the solution in configuration space\ with a different
representation of the Riesz derivative and argued in favor of inconsistency
[14]. To pinpoint the source of this controversy and its resolution, in the
Section IV we scrutinize the different representations of the Riesz derivative
and show that when calculated consistently, they all have the same Fourier
transform. The controversy arises when the divergent integrals in the
configuration space are evaluated piecewise for the infinite square well
problem, thus tampering with the integrity of the Riesz derivative. Finally,
Section V is the conclusions.

\section{Consistency of the Solutions of the Space Fractional Schr\"{o}dinger
Equation}

In one dimension the space fractional Schr\"{o}dinger equation is written in
terms of the quantum Riesz derivative $\left(  -\hslash^{2}\Delta\right)
^{\alpha/2}$ [5-8] as%
\begin{equation}
i\hslash\frac{\partial\Psi(x,t)}{\partial t}=D_{\alpha}\left(  -\hslash
^{2}\Delta\right)  ^{\alpha/2}\Psi(x,t)+V(x)\Psi(x,t),
\end{equation}
where%
\begin{equation}
\left(  -\hslash^{2}\Delta\right)  ^{\alpha/2}\Psi(x,t)=\frac{1}{2\pi\hslash
}\int_{-\infty}^{+\infty}dp\text{ }e^{ipx/\hslash}\left\vert p\right\vert
^{\alpha}\Phi(p,t),\text{ }1<\alpha\leq2,
\end{equation}
and $\Phi(p,t)$ is the Fourier transform of the wave function:
\begin{equation}
\Phi(p,t)=\int_{-\infty}^{+\infty}dx\Psi(x,t)e^{-ipx/\hslash}.
\end{equation}
The restriction on $\alpha$ comes from the requirement of the existence of the
first-order moments of the $\alpha$-stable L\'{e}vy distribution so that
average momentum or position of the quantum particle can be found [5]. For the
infinite square well, the potential is given as
\begin{equation}
V(x)=\left\{
\begin{tabular}
[c]{ccc}%
$0$ & $;$ & $\left\vert x\right\vert <a$\\
$\infty$ & $;$ & $\left\vert x\right\vert \geqslant a$%
\end{tabular}
\ \ \ \ \ \ \right.  ,
\end{equation}
where for its separable solutions:
\begin{equation}
\Psi(x,t)=e^{-iEt/\hslash}\psi(x),
\end{equation}
$\psi(x)$ satisfies the following eigenvalue problem:%
\begin{equation}
D_{\alpha}\left(  -\hslash^{2}\Delta\right)  ^{\alpha/2}\psi(x)=E\psi
(x),\text{ }\psi(a)=\psi(-a)=0.
\end{equation}
The \ corresponding energy eigenfunctions and the eigenvalues are obtained as
[5-8]:%
\begin{align}
\psi_{n}(x)  &  =\left\{
\begin{tabular}
[c]{ccc}%
$A\sin\frac{n\pi}{2a}(x+a)$ & $;$ & $\left\vert x\right\vert <a$\\
&  & \\
$0$ & $;$ & $\left\vert x\right\vert \geqslant a$%
\end{tabular}
\ \ \ \ \ \ \ \right.  ,\\
\text{ }E_{n}  &  =D_{\alpha}\left(  \frac{\hslash n\pi}{2a}\right)  ^{\alpha
},\text{ }n=1,2,\ldots\text{ }.\nonumber
\end{align}

To show the inconsistency of these solutions, Jeng et. al. [9] concentrated on
the ground state with $n=1$:%
\begin{equation}
\psi_{1}(x)=\left\{
\begin{tabular}
[c]{ccc}%
$A\cos\left(  \frac{\pi x}{2a}\right)  $ & $;$ & $\left\vert x\right\vert
<a$\\
&  & \\
$0$ & $;$ & $\left\vert x\right\vert \geqslant a$%
\end{tabular}
\ \ \ \right.
\end{equation}
and argued that this solution, albeit satisfying the boundary conditions,
$\psi_{1}(-a)=\psi_{1}(a)=0,$ when substituted back into the space fractional
Schr\"{o}dinger equation leads to a contradiction [9]. Using the Fourier
transform of $\psi_{1}(x):$
\begin{equation}
\phi_{1}(p)=\mathcal{F}\left\{  \psi_{1}(x)\right\}  =-A\pi\left(
\frac{\hslash^{2}}{a}\right)  \frac{\cos\left(  ap/\hslash\right)  }%
{p^{2}-\left(  \pi\hslash/2a\right)  ^{2}},\text{ }\left\vert x\right\vert <a,
\end{equation}
and the definition of the quantum Riesz derivative [5-8]:%

\begin{equation}
\left(  -\hslash^{2}\Delta\right)  ^{\alpha/2}\psi_{1}(x)=(1/2\pi\hslash
)\int_{-\infty}^{+\infty}dpe^{ipx/\hslash}\left\vert p\right\vert ^{\alpha
}\phi_{1}(p),
\end{equation}
in Equation (6), they wrote $\psi_{1}(x)$ as the integral
\begin{equation}
\psi_{1}(x)=-\frac{AD_{\alpha}}{2E_{1}}\left(  \frac{\hslash}{a}\right)
\int_{-\infty}^{+\infty}dp\text{ }\left(  \frac{2a}{\pi\hslash}\right)
^{2}\frac{\left\vert p\right\vert ^{\alpha}\cos\left(  ap/\hslash\right)
}{\ \left(  2ap/\pi\hslash\right)  ^{2}-1}e^{ipx/\hslash},\text{ }\left\vert
x\right\vert <a.
\end{equation}
Using the substitution $q=\frac{2a}{\pi\hslash}p,$ $\psi_{1}(x)\ $becomes%
\begin{equation}
\psi_{1}(x)=-\frac{AD_{\alpha}}{\pi E_{1}}\left(  \frac{\pi\hslash}%
{2a}\right)  ^{\alpha}\int_{-\infty}^{+\infty}dq\text{ }\frac{\left\vert
q\right\vert ^{\alpha}\cos\left(  \pi q/2\right)  }{\ q^{2}-1}e^{i\pi qx/2a}.
\end{equation}
\ Jeng et. al. [9] argued that the right hand side of the above equation,
which they wrote as%
\begin{equation}
\psi_{1}(x)=-\frac{AD_{\alpha}}{\pi E_{1}}\left(  \frac{\pi\hslash}%
{2a}\right)  ^{\alpha}2\int_{0}^{+\infty}dq\text{ }\frac{\left\vert
q\right\vert ^{\alpha}\cos\left(  \pi q/2\right)  }{\ q^{2}-1}\cos\left(  \pi
qx/2a\right)  ,
\end{equation}
can not satisfy the boundary conditions that $\psi_{1}(x)$ satisfies as
$x\rightarrow\pm a$, thus indicating an inconsistency in the infinite square
well solution. However, we have shown that an exact evaluation of the
integral\ in Equation (12) proves otherwise [10, 11]. In the Section III we
give the general proof for all $n.$

\section{Proof For All $n$}

\subsection{The case for odd $n$}

For\ the odd values of $n,$ eigenfunctions in Equation (7) become%
\begin{align}
\psi_{n}(x)  &  =\left\{
\begin{tabular}
[c]{ccc}%
$A\cos\frac{n\pi x}{2a}$ & $;$ & $\left\vert x\right\vert <a$\\
&  & \\
$0$ & $;$ & $\left\vert x\right\vert \geqslant a$%
\end{tabular}
\ \ \ \ \ \right.  ,\\
\text{ }E_{n}  &  =D_{\alpha}\left(  \frac{\hslash n\pi}{2a}\right)  ^{\alpha
},\text{ }n=1,3,5,\ldots\text{ }.\nonumber
\end{align}
Using the Fourier transform $\mathcal{F}\{\psi_{n}(x)\}=\phi_{n}(p):$%
\begin{equation}
\phi_{n}(p)=-\frac{An\pi\hbar^{2}\sin(n\pi/2)}{a}\left(  \frac{\cos pa/\hbar
}{p^{2}-(n\pi\hbar/2a)^{2}}\right)  ,\text{ }n=1,3,5,\ldots,
\end{equation}
and the definition of the Riesz derivative [Eq. (2)] in the space fractional
Schr\"{o}dinger Equation [Eq. (6)], the corresponding integral expression for
$\psi_{n}(x),$ $n=1,3,5,\ldots$ becomes:%
\begin{equation}
\psi_{n}(x)=-\frac{AD_{\alpha}n\hbar(2a/n\pi\hbar)^{2}\sin(n\pi/2)}{2aE_{n}%
}\int_{-\infty}^{+\infty}dp\text{ }e^{ipx/\hbar}\frac{\left\vert p\right\vert
^{\alpha}\cos(pa/\hbar)}{(2ap/n\pi\hbar)^{2}-1\ }\text{.}%
\end{equation}
Making the substitution $p=(n\pi\hbar/2a)q,$ we write%
\begin{align}
\psi_{n}(x)  &  =-\ \frac{AD_{\alpha}\sin(n\pi/2)}{E_{n}\pi}\ \left(
\frac{n\pi\hbar}{2a}\right)  ^{\alpha}\int_{-\infty}^{+\infty}dq\text{
}e^{i(n\pi x/2a)q}\frac{\left\vert q\right\vert ^{\alpha}\cos(n\pi
q/2)}{\left(  q^{2}-1\right)  \ }\nonumber\\
&  =-\ \frac{AD_{\alpha}\sin(n\pi/2)}{E_{n}\pi}\ \left(  \frac{n\pi\hbar}%
{2a}\right)  ^{\alpha}I,
\end{align}
where $I$ is the integral%
\begin{equation}
I=\int_{-\infty}^{+\infty}dq\text{ }e^{i(n\pi x/2a)q}\frac{\left\vert
q\right\vert ^{\alpha}\cos(n\pi q/2)}{\left(  q^{2}-1\right)  \ }.
\end{equation}
Substituting%
\begin{equation}
\cos\left(  n\pi q/2\right)  =\frac{1}{2}\left(  e^{in\pi q/2}+e^{-in\pi
q/2}\right)  ,
\end{equation}
$\ $we can write $I$ as the sum of two integrals:%
\begin{align}
I  &  =I_{1}+I_{2}\nonumber\\
&  =\frac{1}{2}\int_{-\infty}^{+\infty}dq\text{ }\frac{\left\vert q\right\vert
^{\alpha}\ e^{i(\frac{n\pi x}{2a}+\frac{n\pi}{2})q}}{\ (q+1)(q-1)}\ +\frac
{1}{2}\int_{-\infty}^{+\infty}dq\text{ }\frac{\left\vert q\right\vert
^{\alpha}\ e^{i(\frac{n\pi x}{2a}-\frac{n\pi}{2})q}}{\ (q+1)(q-1)},
\end{align}
which can be evaluated by analytic continuation as a Cauchy principal value
integral [15 pg. 365]. However, In the above integrals, as it stands,
$\left\vert q\right\vert ^{\alpha}$ can not be continued analytically. To
overcome this difficulty, we resort to the original definition of the Riesz
derivative and see where $\left\vert q\right\vert ^{\alpha}$ comes from.

The Riesz derivative, $R_{x}^{\alpha}f(x),$ is defined as [3, 16-18]
\begin{align}
R_{x}^{\alpha}f(x)  &  =-\frac{_{-\infty}D_{x}^{\alpha}f(x)+_{\infty}%
D_{x}^{\alpha}f(x)}{2\cos\alpha\pi/2},\text{ }\alpha>0,\text{ }\alpha
\neq1,3,...\\
_{-\infty}D_{x}^{\alpha}f(x)  &  =\frac{1}{\Gamma(n-\alpha)}\int_{-\infty}%
^{x}(x-x^{\prime})^{-\alpha-1+n}f^{(n)}(x^{\prime})dx^{\prime},\\
_{+\infty}D_{x}^{\alpha}f(x)  &  =\frac{(-1)^{n}}{\Gamma(n-\alpha)}\int
_{x}^{\infty}(x^{\prime}-x)^{-\alpha-1+n}f^{(n)}(x^{\prime})dx^{\prime},
\end{align}
where $n$ is the smallest integer greater than $\alpha.$ For the range
$1<\alpha<2,$ $n=2.$ In Equations (22) and (23) we have used the Caputo
fractional derivative [Eqs. (A7) and (A8)] since for sufficiently smooth
functions:
\begin{equation}
f(x),f^{\prime}(x),\ldots,f^{(n-1)}(x)\rightarrow0\text{ as}~x\rightarrow
\pm\infty,
\end{equation}
the Caputo and the Riemann-Liouville definitions agree [2, 3, 16-18]. Also
note that the quantum Riesz derivative and the Riesz derivative $R_{x}%
^{\alpha}$ are related by [5-8]
\begin{equation}
\left(  -\hslash^{2}\Delta\right)  ^{\alpha/2}\psi_{1}(x)=-\hslash^{\alpha
}R_{x}^{\alpha}\psi_{1}(x).
\end{equation}
Using the following Fourier transforms (see Section IV for the detailed
derivation):%
\begin{equation}
\left.
\begin{tabular}
[c]{l}%
$\mathcal{F}\left\{  _{-\infty}D_{x}^{\alpha}f(x)\right\}  =(i\omega)^{\alpha
}g(\omega),$\\
\\
$\mathcal{F}\left\{  _{\infty}D_{x}^{\alpha}f(x)\right\}  =(-i\omega)^{\alpha
}g(\omega)$%
\end{tabular}
\ \ \ \ \ \ \ \ \right.
\end{equation}
where $g(\omega)=\mathcal{F}\left\{  f(x)\right\}  $ and $\alpha>0$, we write
the Fourier transform of the Riesz derivative as%
\begin{equation}
\mathcal{F}\left\{  R_{x}^{\alpha}f(x)\right\}  =-\left(  \frac{(i\omega
)^{\alpha}+(-i\omega)^{\alpha}}{2\cos\alpha\pi/2}\right)  g(\omega).
\end{equation}
When $\omega$ is restricted to the real axis, this reduces to the familiar
expression $\mathcal{F}\left\{  R_{x}^{\alpha}f(x)\right\}  =-\left\vert
\omega\right\vert ^{\alpha}g(\omega),$ which is used in Equations (12) and
(17). In order to evaluate $I$ by analytic continuation, we use the above form
of the Riesz derivative [Eq. (27)], which allows analytic continuation and
write $I\ $[Eq. (20)] as
\begin{align}
I  &  =I_{1}+I_{2}\nonumber\\
&  =\frac{1}{2}\int_{-\infty}^{+\infty}dq\left(  \frac{(iq)^{\alpha
}+(-iq)^{\alpha}}{2\cos\alpha\pi/2}\right)  \frac{e^{i(\frac{n\pi x}{2a}%
+\frac{n\pi}{2})q}}{\ (q+1)(q-1)}\ \nonumber\\
&  +\frac{1}{2}\int_{-\infty}^{+\infty}dq\left(  \frac{(iq)^{\alpha
}+(-iq)^{\alpha}}{2\cos\alpha\pi/2}\right)  \frac{\ e^{i(\frac{n\pi x}%
{2a}-\frac{n\pi}{2})q}}{\ (q+1)(q-1)}.
\end{align}
Factoring $q^{\alpha}$ out, the integrals $I_{1}$ and $I_{2}$:
\begin{align}
I_{1}  &  =\ \left(  \frac{(i)^{\alpha}+(-i)^{\alpha}}{4\cos\alpha\pi
/2}\right)  \int_{-\infty}^{+\infty}dq\frac{q^{\alpha}e^{i(\frac{n\pi x}%
{2a}+\frac{n\pi}{2})q}}{\ (q+1)(q-1)}\ ,\\
I_{2}  &  =\ \left(  \frac{(i)^{\alpha}+(-i)^{\alpha}}{4\cos\alpha\pi
/2}\right)  \int_{-\infty}^{+\infty}dq\frac{q^{\alpha}e^{i(\frac{n\pi x}%
{2a}-\frac{n\pi}{2})q}}{\ (q+1)(q-1)}\ ,
\end{align}
can now be evaluated as Cauchy principal value integrals via analytic
continuation [15 pg. 365].

In the above integrals, aside from the poles at $q=$ $\pm1,$ there is also a
branch point and a branch cut at the origin due to\ the power $q^{\alpha},$
$\alpha>0.$ For each integral, in contrast to the claims of Hawkins and
Schwarz [12], the branch cut can always be chosen away from the region of
interest. For the branch values of $(i)^{\alpha}$ and $(-i)^{\alpha},$ it has
to be remembered that the Riesz derivative $R_{x}^{\alpha}$, is defined such
that for real $q,$ the Fourier transform of the Riesz derivative corresponds
to the logarithm of the characteristic function of the symmetric L\'{e}vy
probability density function. Therefore, in the definition of the Riesz
derivative [Eq. (21)], $2\cos\alpha\pi/2$ is introduced with the principal
branch values of $(i)^{\alpha}$ and $(-i)^{\alpha}$ in mind, hence
$(i)^{\alpha}+(-i)^{\alpha}=2\cos\alpha\pi/2$. This way, along with the minus
sign introduced by hand in Equation (21), $R_{x}^{\alpha}$ reproduces the
standard derivative $\frac{d^{2}}{dx^{2}}$ for $\alpha=\allowbreak2$ [3, 17].
Other linear combinations of $_{-\infty}D_{x}^{\alpha}f(x)$ and $_{\infty
}D_{x}^{\alpha}f(x)$ have also found use in literature as the Feller
derivative, which gives an additional degree of freedom in terms of a
parameter called the phase or the skewness parameter [3, 17].

\subsubsection{Evaluation of $I_{1}$ and $I_{2}$}

For $I_{1}$ [Eq. (29)] the contour is closed counterclockwise in the upper
half\ complex $q-$plane over a semicircular path with radius $R,$ and then the
contour detours around the poles on the real axis over semicircular paths of
radius $\delta$ in the upper half $q-$plane. Similarly, the contour goes
around the branch point at the origin with the branch cut located in the lower
half of the $q-$plane. Since $\alpha>0$, the integrand vanishes on the contour
as the radius of the semicircular path over $q=0$ shrinks to zero, hence the
integral over the branch point does not contribute to the integral. In the
limit as $R\rightarrow\infty$, by the Jordan's lemma, the contribution coming
from the large semicircle vanishes, thus allowing the evaluation of this
integral as a Cauchy principal value integral in the limit $\delta
\rightarrow0$ as [15 pg. 365]%
\begin{align}
PV(I_{1})  &  =\ \ \left(  \frac{i\pi}{2}\right)  \frac{\sin n\pi/2}%
{4\cos\alpha\pi/2}\left[  (i^{\alpha}+(-i)^{\alpha})(-1+(-1)^{\alpha}%
)\sin\left(  n\pi x/2a\right)  \right. \nonumber\\
&  \left.  +i(i^{\alpha}+(-i)^{\alpha})(1+(-1)^{\alpha})\cos\left(  n\pi
x/2a\right)  \right] \\
&  =-\left(  \frac{\pi\sin n\pi/2}{2}\right)  \cos\left(  n\pi x/2a\right)
,\text{ }n=1,3,\ldots\text{ .}%
\end{align}
Note that one also uses the relation $[i^{\alpha}+(-i)^{\alpha}]=(-1)^{\alpha
}[i^{\alpha}+(-i)^{\alpha}].$

For $I_{2},$ the contour is closed counterclockwise in the lower $q-$plane and
circles around the poles and the branch point in the lower half $q-$plane. For
$I_{2}$ the branch cut is chosen in the upper half $q-$plane and again since
$\alpha>0,$ the integral around the branch point does not contribute to the
integral, thus yielding $PV(I_{2})$ as
\begin{equation}
PV(I_{2})=-\left(  \frac{\pi\sin n\pi/2}{2}\right)  \cos\left(  n\pi
x/2a\right)  ,\text{ }n=1,3,\ldots\text{ ,}%
\end{equation}
which leads to the Cauchy principal value of $I$ as the sum%
\begin{align}
PV(I)  &  =PV(I_{1})+PV(I_{2})\nonumber\\
&  =-\pi\left(  \sin n\pi/2\right)  \cos\left(  n\pi x/2a\right)  ,\text{
}n=1,3,\ldots\text{ .}%
\end{align}
When this is substituted back into Equation (17) we get%

\begin{align}
\psi_{n}(x)  &  =-AD_{\alpha}\frac{\sin(n\pi/2)}{E_{n}\pi}\ \left(  \frac
{n\pi\hbar}{2a}\right)  ^{\alpha}PV(I)\\
&  =\frac{AD_{\alpha}\sin^{2}(\frac{n\pi}{2})}{E_{n}}\ \left(  \frac{n\pi
\hbar}{2a}\right)  ^{\alpha}\cos\frac{n\pi x}{2a}.
\end{align}
Since $E_{n}=D_{\alpha}(\frac{\hslash n\pi}{2a})^{\alpha}$ and $\sin^{2}%
(\frac{n\pi}{2})=1$ for odd $n,$ we again obtain the wave function [Eq. (14)]
as%
\begin{equation}
\psi_{n}(x)=A\cos\frac{n\pi x}{2a},\text{ }n=1,3,\ldots\text{, }\left\vert
x\right\vert <a,
\end{equation}
which on the contrary to Jeng et. al. [9] and Hawkins and Schwarz [12],
vanishes at the boundary as $x\rightarrow\pm a$, hence there is no
inconsistency with the solution outside.

\subsection{The case for even $n$}

The proof for the even $n$ values follows along the same lines [10, 11]. We
first write the wave function [Eq. (7)] as
\begin{align}
\psi_{n}(x)  &  =\left\{
\begin{tabular}
[c]{ccc}%
$A\sin\frac{n\pi x}{2a}$ & $;$ & $\left\vert x\right\vert <a$\\
&  & \\
$0$ & $;$ & $\left\vert x\right\vert \geqslant a$%
\end{tabular}
\ \ \ \ \ \ \ \ \right.  ,\\
\text{ }E_{n}  &  =D_{\alpha}\left(  \frac{\hslash n\pi}{2a}\right)  ^{\alpha
},\text{ }n=2,4,\ldots\text{ },\nonumber
\end{align}
and then obtain its Fourier transform\ as%
\begin{equation}
\phi_{n}(p)=-\frac{iAn\pi\hbar^{2}(\cos n\pi/2)}{a}\frac{\sin\left(
pa/\hbar\right)  }{p^{2}-\left(  n\pi\hbar/2a\right)  ^{2}},\text{
}n=2,4,\ldots\text{ .}%
\end{equation}
Now the integral representation of $\psi_{n}(x)$ becomes
\begin{equation}
\psi_{n}(x)=-\ \frac{iAD_{\alpha}\cos(n\pi/2)}{E_{n}\pi}\ \left(  \frac
{n\pi\hbar}{2a}\right)  ^{\alpha}\int_{-\infty}^{+\infty}dq\text{ }e^{i(n\pi
x/2a)q}\frac{\left\vert q\right\vert ^{\alpha}\sin(n\pi q/2)}{\left(
q^{2}-1\right)  \ },
\end{equation}
where we used the substitution $p=(n\pi\hslash/2a)q.$ Finally, using%
\begin{equation}
\sin\left(  n\pi q/2\right)  =\frac{1}{2i}\left(  e^{in\pi q/2}-e^{-in\pi
q/2}\right)  ,
\end{equation}
and the original definition of the Riesz derivative [Eq.(21)], we write%
\begin{equation}
\psi_{n}(x)=-\ \frac{AD_{\alpha}\cos(n\pi/2)}{E_{n}\pi}\ \left(  \frac
{n\pi\hbar}{2a}\right)  ^{\alpha}I,
\end{equation}
where$\ $%
\begin{align}
I  &  =I_{1}-I_{2},\\
I_{1}  &  =\ \left(  \frac{(i)^{\alpha}+(-i)^{\alpha}}{4\cos\alpha\pi
/2}\right)  \int_{-\infty}^{+\infty}dq\frac{q^{\alpha}e^{i(\frac{n\pi x}%
{2a}+\frac{n\pi}{2})q}}{\ (q+1)(q-1)}\ ,\\
I_{2}  &  =\ \left(  \frac{(i)^{\alpha}+(-i)^{\alpha}}{4\cos\alpha\pi
/2}\right)  \int_{-\infty}^{+\infty}dq\frac{q^{\alpha}e^{i(\frac{n\pi x}%
{2a}-\frac{n\pi}{2})q}}{\ (q+1)(q-1)}\ .
\end{align}
The Cauchy principal value of $I$ is now found as%
\begin{equation}
PV(I)=-\pi\left(  \cos n\pi/2\right)  \sin\left(  n\pi x/2a\right)  ,\text{
}n=2,4,\ldots\text{ },
\end{equation}
which when substituted into (42) yields the wave function in (38),
hence\ again no inconsistency.

This is not surprising at all. In fact, Equations (14) and (17) and similarly
Equations (38) and (40), represent the same wave function, where Equations
(17) and (40) are just the integral representations of $\psi_{n}(x)$ in
Equations (14) and (38) for the odd and the even values of $n$, respectively.
It is true that the Riesz derivative is a non local operator [Eqs. (21-23)]
that requires knowledge of the wave function over the entire space. For the
infinite square well problem, the system is confined to the region $\left\vert
x\right\vert <a$ with \ $\Psi(x,t)=0$ for $\left\vert x\right\vert \geq a.$
Since the solution for $\left\vert x\right\vert <a$ satisfies the boundary
conditions as $x\rightarrow\pm a,$ the solution inside the well is consistent
with the outside.

\section{Scrutinizing the Riesz Derivative}

Another source for the proposed inconsistency in the infinite square well
solution [Eq. (7)] is that when the Riesz derivative in Equation (21) is
directly calculated by evaluating the integrals in Equations (22) and (23),
the result does not satisfy the space fractional Schr\"{o}dinger equation
[14]. Note that these integrals are now in the configuration space. This
situation is explained by the fact that the Riesz derivative is non local,
hence to find the solution outside the well, one also has to consider the
solution inside [14]. To shed some light on this problem, we now scrutinize
how the different definitions of the Riesz derivative are written and how they
are\ related and calculated.

\subsection{Riesz Fractional Integral}

To evaluate the integrals in the definition of $R_{x}^{\alpha}f(x)$ [Eqs. (22)
and (23)], we are going to start with the definition of the Riesz fractional
integral, which is defined as [Eqs. (A1) and (A2)]%
\begin{align}
R_{x}^{-\alpha}f(x)  &  =\frac{_{-\infty}D_{x}^{-\alpha}f(x)+_{\infty}%
D_{x}^{-\alpha}f(x)}{2\cos\alpha\pi/2},\text{ }\ \alpha>0,\text{ }\alpha
\neq1,3,...,\\
_{-\infty}D_{x}^{-\alpha}f(x)  &  =\frac{1}{\Gamma(\alpha)}\int_{-\infty}%
^{x}(x-x^{\prime})^{\alpha-1}f(x^{\prime})dx^{\prime},\\
_{+\infty}D_{x}^{-\alpha}f(x)  &  =\frac{1}{\Gamma(\alpha)}\int_{x}^{\infty
}(x^{\prime}-x)^{\alpha-1}f(x^{\prime})dx^{\prime}.
\end{align}
To evaluate the integral in Equation (48), we define the function
\begin{equation}
h_{+}(x)=\left\{
\begin{array}
[c]{ccc}%
\frac{x^{\alpha-1}}{\Gamma(\alpha)} & , & x>0\\
&  & \\
0 & , & x\leq0
\end{array}
\right.  ,
\end{equation}
which allows us to write $_{-\infty}D_{x}^{-\alpha}f(x)$ as the convolution of
$h_{+}(x)$ with $f(x)$:%
\begin{equation}
_{-\infty}D_{x}^{-\alpha}f(x)=h_{+}(x)\ast f(x).
\end{equation}
It is well known that the Fourier transform of a convolution is equal to the
product of the Fourier transforms of the convolved functions, that is,%
\begin{equation}
\mathcal{F}\left\{  _{-\infty}D_{x}^{-\alpha}f(x)\right\}  =\mathcal{F}%
\left\{  h_{+}(x)\right\}  \mathcal{F}\left\{  f(x)\right\}  .
\end{equation}
Using analytic continuation with an appropriate contour, it is straight
forward to evaluate the Fourier transform of $h_{+}(x)$ as
\begin{equation}
\mathcal{F}\left\{  h_{+}(x)\right\}  =\int_{-\infty}^{\infty}\frac
{x^{\alpha-1}}{\Gamma(\alpha)}e^{-i\omega x}dx=(i\omega)^{-\alpha},\text{
}\alpha>0.
\end{equation}
Assuming that the Fourier transform of $f(x)$ exists:
\begin{equation}
\mathcal{F}\left\{  f(x)\right\}  =\int_{-\infty}^{\infty}f(x)e^{-i\omega
x}dx=F(\omega),
\end{equation}
which only demands an absolutely integrable $f(x)$, we obtain the Fourier
transform%
\begin{equation}
\mathcal{F}\left\{  _{-\infty}D_{x}^{-\alpha}f(x)\right\}  =(i\omega
)^{-\alpha}F(\omega),\text{ }\alpha>0.
\end{equation}
Following similar steps, we define the function%
\begin{equation}
h_{-}(x)=\left\{
\begin{array}
[c]{ccc}%
0 & , & x\geq0\\
&  & \\
\frac{(-x)^{\alpha-1}}{\Gamma(\alpha)} & , & x<0
\end{array}
\right.  ,
\end{equation}
with the Fourier transform%
\begin{equation}
\mathcal{F}\left\{  h_{-}(x)\right\}  =\int_{-\infty}^{0}\frac{(-x)^{\alpha
-1}}{\Gamma(\alpha)}e^{-i\omega x}dx=(-i\omega)^{-\alpha},\text{ }\alpha>0.
\end{equation}

We can now write $_{\infty}D_{x}^{-\alpha}f(x)$ [Eq. (49)] as the convolution%
\begin{equation}
_{\infty}D_{x}^{-\alpha}f(x)=h_{-}(x)\ast f(x),
\end{equation}
where its Fourier transform is given as%
\begin{align}
\mathcal{F}\left\{  _{\infty}D_{x}^{-\alpha}f(x)\right\}   &  =\mathcal{F}%
\left\{  h_{-}(x)\right\}  \mathcal{F}\left\{  f(x)\right\} \\
&  =(-i\omega)^{-\alpha}F(\omega),\text{ }\alpha>0.
\end{align}
Using Equations (55) and (60), the Riesz fractional integral, $R_{x}^{-\alpha
}f(x),$ is defined in terms of its Fourier transform as%
\begin{align}
\mathcal{F}\left\{  R_{x}^{-\alpha}f(x)\right\}   &  =\frac{(i\omega
)^{-\alpha}+(-i\omega)^{-\alpha}}{2\cos\alpha\pi/2}F(\omega),\text{ }%
\alpha>0,\text{ }\alpha\neq1,3,...,\\
&  =\left\vert \omega\right\vert ^{-\alpha}F(\omega),\text{ for real }\omega.
\end{align}
Also note that from Equations (47-49),~$R_{x}^{-\alpha}f(x)$ is also the
integral%
\begin{equation}
R_{x}^{-\alpha}f(x)=\frac{1}{2\Gamma(\alpha)\cos\alpha\pi/2}\int_{-\infty
}^{\infty}\left\vert x-x^{\prime}\right\vert ^{\alpha-1}f(x^{\prime
})dx^{\prime},\text{ }\alpha>0,\text{ }\alpha\neq1,3,...\text{ }.
\end{equation}

\subsection{Riesz Fractional Derivative}

To evaluate the Riesz fractional derivative, we note that in Equations
(21-23), the Caputo definition of the fractional derivative is used. Since for
sufficiently smooth functions [Eq. (24)]:
\[
f(x),f^{\prime}(x),\ldots,f^{(n-1)}(x)\rightarrow0\text{ as}~x\rightarrow
\pm\infty,
\]
the Caputo and the Riemann-Liouville definitions agree [Eq. (A9)], we can
write Equation (22) as [Eq. (A7) [3, 16-18]]%
\begin{align}
_{-\infty}D_{x}^{\alpha}f(x) &  =\frac{1}{\Gamma(n-\alpha)}\int_{-\infty}%
^{x}(x-x^{\prime})^{-\alpha-1+n}f^{(n)}(x^{\prime})dx^{\prime},\text{ }%
\alpha>0,\\
&  =_{-\infty}\mathbf{I}_{x}^{n-\alpha}f^{(n)}(x)=_{-\infty}D_{x}^{\alpha
-n}f^{(n)}(x).
\end{align}
Note that we have dropped the abbreviation $R-L$ and $C$ in $^{R-L}%
D_{x}^{\alpha}$ and $^{C}D_{x}^{\alpha}$. Since $\alpha-n<0,$ we can use our
previous result [Eq. (55)] to obtain [2, 16]%
\begin{align}
\mathcal{F}\left\{  _{-\infty}D_{x}^{\alpha}f(x)\right\}   &  =\mathcal{F}%
\left\{  _{-\infty}D_{x}^{\alpha-n}f^{(n)}(x)\right\}  \\
&  =(i\omega)^{\alpha-n}\mathcal{F}\left\{  f^{(n)}(x)\right\}  \\
&  =(i\omega)^{\alpha-n}(i\omega)^{n}F(\omega)\\
&  =(i\omega)^{\alpha}F(\omega).
\end{align}
The third step [Eq. (68)], is already assured by the smoothness condition [Eq.
(24)]. Similarly, we obtain%
\begin{equation}
\mathcal{F}\left\{  _{\infty}D_{x}^{\alpha}f(x)\right\}  =(-i\omega)^{\alpha
}F(\omega).
\end{equation}
Therefore, we can write the Fourier transform of the Riesz derivative [Eq.
(21)] as%
\begin{equation}
\mathcal{F}\left\{  R_{x}^{\alpha}f(x)\right\}  =-\frac{(i\omega)^{\alpha
}+(-i\omega)^{\alpha}}{2\cos\alpha\pi/2}F(\omega),\text{ }\alpha>0,\text{
}\alpha\neq1,3,...,
\end{equation}
where $F(\omega)$ is the Fourier transform of $f(x)$ [Eq. (54)], which makes
use of the values of $f(x)$ over the entire range $x\in(-\infty,\infty)$. For
real $\omega,$ we can also write this as%
\begin{equation}
\mathcal{F}\left\{  R_{x}^{\alpha}f(x)\right\}  =-\left\vert \omega\right\vert
^{\alpha}F(\omega),
\end{equation}
which was used to write Equations (12), (17) and (40). So far, all we have
assumed is that $f(x)$ is absolutely integrable, hence its Fourier transform
exists and the smoothness condition in Equation (24). Granted that the inverse
transform exists, the Riesz derivative is defined as
\begin{align}
R_{x}^{\alpha}f(x) &  =\mathcal{F}^{-1}\left\{  -\left\vert \omega\right\vert
^{\alpha}F(\omega)\right\}  \\
&  =-\frac{1}{2\pi}\int_{-\infty}^{\infty}\left\vert \omega\right\vert
^{\alpha}F(\omega)e^{i\omega x}d\omega.
\end{align}

Note that our starting point was the integrals in Equations (22$-$23), hence
using (21), $R_{x}^{\alpha}f(x)$ can also be written as%
\begin{gather}
R_{x}^{\alpha}f(x)\ =-\frac{1}{2\Gamma(2-\alpha)\cos\alpha\pi/2}\\
\times\left[  \int_{-\infty}^{x}(x-x^{\prime})^{-\alpha+1}f^{(2)}(x^{\prime
})dx^{\prime}+\int_{x}^{\infty}(x^{\prime}-x)^{-\alpha+1}f^{(2)}(x^{\prime
})dx^{\prime}\right]  ,\text{ }1<\alpha<2,\nonumber
\end{gather}
where we have set $n=2$ for $1<\alpha<2.$

It is important to note that Equations (74) and (75) correspond to different
representations of the Riesz derivative, which have the same Fourier
transform. As we have shown, Equation (74) is actually obtained from the
Fourier transform of (75). It is not true to say that non local effects are
incorporated in (75) but not in (74).\ In Equation (74),\ the Fourier
transform of $f(x)$ is obtained by integrating over the entire space as
$F(\omega)=\int_{-\infty}^{\infty}f(x)e^{-i\omega x}dx.$ In (74),
$R_{x}^{\alpha}f(x)$ is given in terms of an integral in the frequency
(momentum) space, while in (75), $R_{x}^{\alpha}f(x)$ is given in terms of
integrals in the configuration space. In general, the integrals in both of
these expressions are singular in their respective spaces. Granted that these
singular integrals\ are treated consistently, they should yield the same
result. However, technically, it is easier to work in the momentum space with
Equation (74).

\subsection{Riesz Derivative via the R-L Definition}

In Equation (75) we have used the Caputo fractional derivative for $_{-\infty
}D_{x}^{\alpha}f(x)$ and $_{\infty}D_{x}^{\alpha}f(x)$. If we use the
Riemann-Liouville definition, The Riesz derivative [Eqs. (21$-$23)] becomes
[Eqs. (A4) and (A6), [3, 16$-$18]]%
\begin{align}
R_{x}^{\alpha}f(x)  &  =-\frac{_{-\infty}D_{x}^{\alpha}f(x)+_{\infty}%
D_{x}^{\alpha}f(x)}{2\cos\alpha\pi/2},\text{ }\alpha>0,\text{ }\alpha
\neq1,3,...,\\
_{-\infty}D_{x}^{\alpha}f(x)  &  =\frac{1}{\Gamma(2-\alpha)}\frac{d^{2}%
}{dx^{2}}\int_{-\infty}^{x}(x-x^{\prime})^{-\alpha+1}f(x^{\prime})dx^{\prime
},\\
_{+\infty}D_{x}^{\alpha}f(x)  &  =\frac{1}{\Gamma(2-\alpha)}\frac{d^{2}%
}{dx^{2}}\int_{x}^{\infty}(x^{\prime}-x)^{-\alpha+1}f(x^{\prime})dx^{\prime},
\end{align}
hence we can also write%
\begin{gather}
R_{x}^{\alpha}f(x)\ =-\frac{1}{2\Gamma(2-\alpha)\cos\alpha\pi/2}\\
\times\left[  \frac{d^{2}}{dx^{2}}\int_{-\infty}^{x}(x-x^{\prime})^{-\alpha
+1}f(x^{\prime})dx^{\prime}+\frac{d^{2}}{dx^{2}}\int_{x}^{\infty}(x^{\prime
}-x)^{-\alpha+1}f(x^{\prime})dx^{\prime}\right]  ,\nonumber
\end{gather}
Using the functions $h_{\pm}(x)$ [Eqs. (50) and (56)] and the convolution
theorem, it is straight forward to show that the Fourier transform of (79) is
still given by Equation (71), or (72) when $\omega$ is real.

\subsection{Source of the Controversy}

The so called inconsistency problem of the infinite square well, in the
configuration space [14] originates from the piecewise evaluation of the
highly singular integrals in Equation (79), which tampers with the integrity
of the Riesz derivative, thus affecting its Fourier transform. For example,
for a point outside the well, say $x\geq a,$ if we write the Riesz derivative
[Eq (79)] as%
\begin{gather}
R_{x}^{\alpha}\psi_{n}(x)\ =-\frac{1}{2\Gamma(2-\alpha)\cos\alpha\pi
/2}\nonumber\\
\times\left\{  \left[  \frac{d^{2}}{dx^{2}}\int_{-\infty}^{-a}(x-x^{\prime
})^{-\alpha+1}\psi_{n}(x^{\prime})dx^{\prime}+\frac{d^{2}}{dx^{2}}\int
_{-a}^{a}(x-x^{\prime})^{-\alpha+1}\psi_{n}(x^{\prime})dx^{\prime}\right.
\right. \nonumber\\
\left.  +\frac{d^{2}}{dx^{2}}\int_{a}^{x}(x-x^{\prime})^{-\alpha+1}\psi
_{n}(x^{\prime})dx^{\prime}\right] \nonumber\\
+\left.  \left[  \frac{d^{2}}{dx^{2}}\int_{x}^{\infty}(x^{\prime}%
-x)^{-\alpha+1}\psi_{n}(x^{\prime})dx\right]  \right\}  ,\text{ }1<\alpha<2~,
\end{gather}
and then substitute the square well solution [Eq. (7)], we obtain%
\begin{equation}
R_{x}^{\alpha}\psi_{n}(x)\ =-\frac{1}{2\Gamma(2-\alpha)\cos\alpha\pi/2}\left[
\frac{d^{2}}{dx^{2}}\int_{-a}^{a}(x-x^{\prime})^{-\alpha+1}\psi_{n}(x^{\prime
})dx^{\prime}\right]  ,\text{ }x\geq a.
\end{equation}
The above expression gives the values of the Riesz derivative outside the
well, $x\geq a$, in terms of an integral that only makes use of the values of
the wave function inside the well. In general, the $R_{x}^{\alpha}\psi_{n}(x)$
given above for $x\geq a$ does not vanish, hence does not satisfy the space
fractional Schr\"{o}dinger equation [Eq. (6)] for $x\geq a$. This implies a
potential problem for the infinite square well solution [14]. Note that to
write Equation (81), we have used the fact that the wave function outside is
zero. Thus, along with the first and the third integrals in Equation (80), we
have set the last integral to zero [14]. Even though this procedure looks
reasonable, what it essentially does is to set the fractional derivative
$_{\infty}D_{x}^{\alpha}\psi_{n}(x)$ to zero for $x\geq a$, that is,
\begin{align}
_{\infty}D_{x}^{\alpha}\psi_{n}(x)\  &  =\frac{1}{\Gamma(2-\alpha)}\frac
{d^{2}}{dx^{2}}\int_{x\geq a}^{\infty}(x^{\prime}-x)^{-\alpha+1}\psi
_{n}(x^{\prime})dx^{\prime}\\
&  =0,\text{ }x\geq a,
\end{align}
thus the Fourier transform $\mathcal{F}\left\{  _{\infty}D_{x}^{\alpha}%
\psi_{n}(x)\right\}  $ is also set to zero for $x\geq a$. However, in the
definition of the Riesz derivative [Eqs. (21$-$23)], the Fourier transform of
$_{\infty}D_{x}^{\alpha}\psi_{n}(x),$ for all $x,$ is given as [Eq. (70)]
\begin{equation}
\mathcal{F}\left\{  _{\infty}D_{x}^{\alpha}\psi_{n}(x)\ \right\}
=(-i\omega)^{\alpha}\Phi_{n}(\omega),
\end{equation}
where $\Phi_{n}(\omega)$ is the Fourier transform of the entire solution,
$\psi_{n}(x)$, not just the solution for\ $x\geq0$.

Similarly, this procedure also tampers with the Fourier transform of
$_{-\infty}D_{x}^{\alpha}\psi_{n}(x)$, thus the Fourier transform of the
derivative in (81) is not what it should be, that is, $\mathcal{F}\left\{
R_{x}^{\alpha}f(x)\right\}  =-\left\vert \omega\right\vert ^{\alpha}%
F(\omega),$ which is the basic definition of the Riesz derivative used in the
space fractional Schr\"{o}dinger equation.

Similarly, the expressions for $x\leq-a$ and $\left\vert x\right\vert <a$ can
be written as [14]%
\begin{align}
R_{x}^{\alpha}\psi_{n}(x)\  &  =-\frac{1}{2\Gamma(2-\alpha)\cos\alpha\pi
/2}\left[  \frac{d^{2}}{dx^{2}}\int_{-a}^{a}(x^{\prime}-x)^{-\alpha+1}\psi
_{n}(x^{\prime})dx^{\prime}\right]  ,\text{ }x\leq-a,\\
R_{x}^{\alpha}\psi_{n}(x)\  &  =-\frac{1}{2\Gamma(2-\alpha)\cos\alpha\pi
/2}\left[  \frac{d^{2}}{dx^{2}}\int_{-a}^{a}\left\vert x-x^{\prime}\right\vert
^{-\alpha+1}\psi_{n}(x^{\prime})dx^{\prime}\right]  ,\text{ }\left\vert
x\right\vert <a.
\end{align}
Note that Equation (81) can also be written as
\begin{align}
R_{x}^{\alpha}\psi_{n}(x)  &  =-\frac{1}{2\Gamma(2-\alpha)\cos\alpha\pi/2}%
\int_{-a}^{a}\frac{\psi_{n}(x^{\prime})}{(x-x^{\prime})^{\alpha+1}}dx^{\prime
},\text{ }x\geq a,\\
&  =-\frac{(-\alpha+1)(-\alpha)}{2\Gamma(2-\alpha)\cos\alpha\pi/2}\int
_{-a}^{a}\frac{\psi_{n}(x^{\prime})}{(x-x^{\prime})^{\alpha+1}}dx^{\prime},\\
&  =-\frac{1}{2\Gamma(-\alpha)\cos\alpha\pi/2}\int_{-a}^{a}\frac{\psi
_{n}(x^{\prime})}{(x-x^{\prime})^{\alpha+1}}dx^{\prime},
\end{align}
This result was used in [14~], which was obtained by using another
representation of the Riesz derivative:%
\begin{equation}
R_{x}^{\alpha}f(x)=\ \frac{\Gamma(1+\alpha)\sin\alpha\pi/2}{\pi}\ \int
_{0}^{\infty}\frac{f(x+x^{\prime})-2f(x)+f(x-x^{\prime})}{x^{\prime\alpha+1}%
}dx^{\prime},
\end{equation}
which is also good for $\alpha=1.$ This representation is obtained by writing
$_{-\infty}D_{x}^{\alpha}f(x)$ and $_{\infty}D_{x}^{\alpha}f(x)$ in Equations
(77) and (78) as [3]%
\begin{align}
_{-\infty}D_{x}^{\alpha}f(x)  &  =\frac{\alpha}{\Gamma(1-\alpha)}\int
_{0}^{\infty}\frac{f(x)-f(x-x^{\prime})}{x^{\prime\alpha+1}}dx^{\prime},\\
_{\infty}D_{x}^{\alpha}f(x)  &  =-\frac{\alpha}{\Gamma(1-\alpha)}\int
_{0}^{\infty}\frac{f(x+x^{\prime})-f(x)}{x^{\prime\alpha+1}}dx^{\prime}.
\end{align}
Similarly for $x\leq-a$ $\ $and $\left\vert x\right\vert <a,$ we obtain the
expressions used in [14] as%
\begin{align}
R_{x}^{\alpha}\psi_{n}(x)\  &  =-\frac{1}{2\Gamma(-\alpha)\cos\alpha\pi
/2}\left[  \int_{-a}^{a}\frac{\psi_{n}(x^{\prime})}{(x^{\prime}-x)^{\alpha+1}%
}dx^{\prime}\right]  ,\text{ }x\leq-a,\\
R_{x}^{\alpha}\psi_{n}(x)\  &  =-\frac{1}{2\Gamma(2-\alpha)\cos\alpha\pi
/2}\left[  \frac{d^{2}}{dx^{2}}\int_{-a}^{a}\left\vert x-x^{\prime}\right\vert
^{-\alpha+1}\psi_{n}(x^{\prime})dx^{\prime}\right]  ,\text{ }\left\vert
x\right\vert <a.
\end{align}

In summary, the Riesz derivative can be evaluated by using Equation (74),
which involves an integration in the frequency (momentum) space. We have shown
that for the infinite square well problem, the use of Equation (74) gives
consistent results. We can also use the representations in Equations (75) or
(79), which involve integrals in configuration space. What is important is
that a consistent treatment of all the representations of the Riesz derivative
should yield the same Fourier transform, that is, $\mathcal{F}\left\{
R_{x}^{\alpha}f(x)\right\}  =-\left\vert \omega\right\vert ^{\alpha}%
F(\omega).$

\section{Conclusions}

Using the convolution theorem we have demonstrated how the frequency
(momentum) space representation of the Riesz derivative [Eq. (74)]:
\begin{equation}
R_{x}^{\alpha}f(x)=-\frac{1}{2\pi}\int_{-\infty}^{\infty}\left\vert
\omega\right\vert ^{\alpha}F(\omega)e^{i\omega x}d\omega,
\end{equation}
is obtained from the integral representations in the configuration space [Eq.
(75)]:%
\begin{gather}
R_{x}^{\alpha}f(x)\ =-\frac{1}{2\Gamma(2-\alpha)\cos\alpha\pi/2}\\
\times\left[  \int_{-\infty}^{x}(x-x^{\prime})^{-\alpha+1}f^{(2)}(x^{\prime
})dx^{\prime}+\int_{x}^{\infty}(x^{\prime}-x)^{-\alpha+1}f^{(2)}(x^{\prime
})dx^{\prime}\right]  ,\text{ }1<\alpha<2,\nonumber
\end{gather}
and similarly from [Eq. (79)]%
\begin{gather}
R_{x}^{\alpha}f(x)\ =-\frac{1}{2\Gamma(2-\alpha)\cos\alpha\pi/2}\\
\times\left[  \frac{d^{2}}{dx^{2}}\int_{-\infty}^{x}(x-x^{\prime})^{-\alpha
+1}f(x^{\prime})dx^{\prime}+\frac{d^{2}}{dx^{2}}\int_{x}^{\infty}(x^{\prime
}-x)^{-\alpha+1}f(x^{\prime})dx^{\prime}\right]  ,\text{ }1<\alpha<2.\nonumber
\end{gather}
Granted that $f(x)$ is absolutely integrable and the smoothness condition in
Equation (24) is satisfied, all the above representations of the Riesz
derivative agree and have the same Fourier transform. The first definition is
given in the frequency (momentum) domain while the others are in the
configuration space.

For the infinite square well, the controversy proposed in [9, 12] is based on
the use of the momentum space definition in Equation (95). In Section II and
III, we have shown that if the relevant integrals are evaluated as Cauchy
principal value integrals, there is no inconsistency.

As for the inconsistency of the infinite well solution proposed in terms of
the configuration space definitions of the Riesz derivative [14], the
segmented evaluation of these integrals leads to the Riesz derivative in
Equation (81) for $x\geq a$, (85) for $x\leq-a$ and (86) for $\left\vert
x\right\vert <a$. Substituting the eigenfunctions [Eq. (7)] into Equations
(89), (93) and (94) we obtain%
\begin{align}
R_{x}^{\alpha}\psi_{n}(x)  &  =F_{1}(x)=-\frac{1}{2\Gamma(-\alpha)\cos
\alpha\pi/2}\int_{-a}^{a}\frac{\psi_{n}(x^{\prime})}{(x-x^{\prime})^{\alpha
+1}}\\
&  =-\frac{A}{2\Gamma(-\alpha)\cos\alpha\pi/2}\int_{-a}^{a}\frac{\sin
\frac{n\pi}{2a}(x^{\prime}+a)}{(x-x^{\prime})^{\alpha+1}},\text{ }x\geq a
\end{align}
and%
\begin{align}
R_{x}^{\alpha}\psi_{n}(x)  &  =F_{2}(x)=-\frac{A}{2\Gamma(-\alpha)\cos
\alpha\pi/2}\left[  \int_{-a}^{a}\frac{\sin\frac{n\pi}{2a}(x^{\prime}%
+a)}{(x^{\prime}-x)^{\alpha+1}}dx^{\prime}\right]  ,\text{ }x\leq-a,\\
R_{x}^{\alpha}\psi_{n}(x)  &  =F_{3}(x)=-\frac{A}{2\Gamma(2-\alpha)\cos
\alpha\pi/2}\left[  \frac{d^{2}}{dx^{2}}\int_{-a}^{a}\frac{\sin\frac{n\pi}%
{2a}(x^{\prime}+a)}{\left\vert x-x^{\prime}\right\vert ^{\alpha-1}}dx^{\prime
}\right]  ,\text{ }\left\vert x\right\vert <a,
\end{align}
which are used in [14] to argue for inconsistency. In these expressions,
$F_{1}(x),$ $F_{2}(x)$ and $F_{3}(x)$ are functions of $x,$ in their
respective intervals. However, since all the integrands are singular at the
end points, none of these functions are well defined, thus\ the integrals do
not exist in the Riemann sense. In this regard, their Fourier transforms do
not exist. The segmented evaluation of the integrals destroys the wholeness in
the definition of the Riesz derivative, hence does not yield the correct
Fourier transform.

In other words, what the above procedure yields in Equations (99-101) is not
the Riesz derivative used in the space fractional Schr\"{o}dinger equation. It
does not have the correct Fourier transform. It has to be kept in mind that
the Riesz derivative is basically defined in terms of its Fourier transform,
which is equal to the logarithm of the characteristic function of the L\'{e}vy
probability distribution function. This is in keeping with one of the basic
premises of the quantum mechanics, which says that the wave functions in
position and momentum spaces are related to each other through a Fourier
transform. This also shows in the fact that the space fractional
Schr\"{o}dinger equation follows from the Feynman path integral formulation of
quantum mechanics over L\'{e}vy paths.

\appendix

\section{Basic Definitions of the Fractional Derivatives and Integrals}

The right- and the left- handed Riemann-Liouville integrals, are defined,
respectively, as [ 2, 3, 16-18]%
\begin{align}
_{a^{+}}\mathbf{I}_{x}^{q}[f(x)]  &  =\frac{1}{\Gamma(q)}\int_{a}^{x}%
(x-\tau)^{q-1}f(\tau)d\tau,\\
_{b^{-}}\mathbf{I}_{x}^{q}[f(x)]  &  =\frac{1}{\Gamma(q)}\int_{x}^{b}%
(\tau-x)^{q-1}f(\tau)d\tau,
\end{align}
where $a<x<b$ and $q>0.$ In applications we frequently encounter cases with
$a=-\infty$ or $b=\infty.$ Fractional integrals with either the lower or the
upper limit is taken as infinity are also called the Weyl fractional integral.
Some authors may reverse the definitions of the right- and the left- handed
derivatives. Sometimes $_{a^{+}}\mathbf{I}_{x}^{q}$ and $_{b^{-}}%
\mathbf{I}_{x}^{q}$ are also called progressive and regressive, respectively.

The right- and the left- handed Riemann-Liouville derivatives of order $q>0$
are defined as [2, 3, 16-18]%
\begin{align}
_{a^{+}}D_{x}^{q}f(x)  &  =\frac{d^{n}}{dx^{n}}\text{ }\left(  _{a^{+}%
}\mathbf{I}_{x}^{n-q}[f(x)]\right) \\
&  =\frac{1}{\Gamma(n-q)}\frac{d^{n}}{dx^{n}}\int_{a}^{x}(x-\tau
)^{n-q-1}f(\tau)d\tau,\\
_{b^{-}}D_{x}^{q}f(x)  &  =(-1)^{n}\frac{d^{n}}{dx^{n}}\text{ }\left(
_{b^{-}}\mathbf{I}_{x}^{n-q}[f(x)]\right) \\
&  =\frac{(-1)^{n}}{\Gamma(n-q)}\frac{d^{n}}{dx^{n}}\int_{t}^{b}%
(\tau-x)^{n-q-1}f(\tau)d\tau,
\end{align}
where $a<x<b$ and $n>q.$

The right-handed Caputo derivative for $q>0$ is defined as%
\begin{align}
_{a^{+}}^{C}D_{x}^{q}f(x)  &  =\text{ }_{a^{+}}\mathbf{I}_{x}^{n-q}%
f^{(n)}(x)\nonumber\\
&  =\frac{1}{\Gamma(n-q)}\int_{a}^{x}\frac{f^{(n)}(\tau)d\tau}{(x-\tau
)^{1-n+q}},
\end{align}
where $n$ is the next integer higher than $q.$

\bigskip The left-handed Caputo derivative for $q>0$ is defined as [2, 3,
16-18]%
\begin{align}
_{b^{-}}^{C}D_{x}^{q}f(x)  &  =(-1)^{n}{}_{b^{-}}\mathbf{I}_{x}^{n-q}%
f^{(n)}(x)\nonumber\\
&  =\frac{(-1)^{n}}{\Gamma(n-q)}\int_{x}^{b}\frac{f^{(n)}(\tau)d\tau}%
{(\tau-x)^{1-n+q}},
\end{align}
where $n$ is again the next integer higher than $q$ [16-18]. We reserve the
letter $a$ for the lower limit of the integral operators and the letter $b$
for the upper limit, hence we will ignore the superscripts in $a^{+}$ and
$b^{-}.$

The two derivatives are related by \ %

\begin{equation}
_{0}^{C}D_{x}^{q}f(x)=\text{ }_{0}^{R-L}D_{x}^{q}f(x)-\sum_{k=0}^{n-1}%
\frac{x^{k-q}}{\Gamma(k-q+1)}f^{(k)}(a^{+}),\text{ }q>0,\text{ }n-1<q<n.
\end{equation}
In other words, the two derivatives are equal when $f(x)$ and its first $n-1$
derivatives vanish at $x=a$.

\end{document}